\def\be{\begin{equation}}
\def\ee{\end{equation}}
\def\bea{\begin{eqnarray}}

\def\eea{\end{eqnarray}}

\def\o{\omega}

\def\s{\sigma}

\def\bd{{\bf d}}

\def\bk{{\bf k}}
\def\bx{{\bf x}}
\def\bX{{\bf X}}

\def\ha{{1\over 2}}

\newskip\humongous \humongous=0pt plus 1000pt minus 1000pt

\newif\ifdtup

\def\(#1){(\ref{#1})}

\documentstyle[12pt]{article}

\makeatletter                    
\@addtoreset{equation}{section}  
\makeatother                     

\begin{document}

\title{Decoherence of Two-Level Systems Can Be Very Different from Brownian Particles}
\author{ B. L. Hu \thanks{hub@physics.umd.edu}\\
{\small Department of Physics, University of Maryland,
 College Park, Maryland 20742}}
\date{\small ({\it umdpp 02-038, Feb. 27, 2002}.
To appear in {\sl Chaos, Solitons, Fractals})} \maketitle

\begin{abstract}

In quantum computation, it is of paramount importance to locate
the parameter space where maximal coherence can be preserved in
the qubit system. Considerable insight in  environment-induced
decoherence has been gained in the last decade from detailed
studies using the quantum Brownian motion (QBM) models.  A number
of respectable authors have applied these insights derived from
QBM models to two level systems interacting with a field. Their
conclusions based on this particular type of qubit coupling to
the environment had led to the general belief that 2LS are easily
decohered.  In a recent paper \cite{AH},  we debunk such a myth
and caution indiscriminate application of the QBM model of
decoherence to arbitrary two level systems. We point out that at
least for a  two-level atom (2LA)- electromagnetic field (EMF)
system alone, as used in the atom cavity prototypes of quantum
computers, the decoherence time is rather long, comparable to the
relaxation time. In the standard Hamiltonian of the 2LA, the
dominant interaction  is the $\hat \sigma_{\pm}$ type of coupling
between the two levels (what constitutes the qubit) and the
field, not  the  $\hat \sigma_z$ type assumed in most previous
discussions of qubit decoherence, which shows the QBM behavior.
Depending on the coupling the field can act as a resonator (in an
atom cavity) or as a bath (in QBM) and produce very different
decoherent behavior. Our conclusion is based on a new exact
master equation we derived at zero temperature which generalizes
the text-book ones restricted by the Born-Markov approximation.
Indeed many cavity experiments testify to the correctness of
these results, and that the 2LA-EMF system maintaining its
coherence in sufficiently long duration is the reason why
experimentalists can manipulate them to show interesting quantum
coherence effects.
\end{abstract}


\section{Introduction}

A two-level system (2LS) interacting with an electromagnetic
field (EMF) has proven to be a very useful model for a wide range
of problems from  atomic-optical \cite{WM,MW,Scu,Wei,VW,CPP,Car}
and condensed matter \cite{Leg87,Weiss} processes to quantum
computation \cite{QComp}. For the latter application stringent
limits in maintaining the coherence of  the 2LS (called qubits)
are required. This prompted us  to revisit the theoretical
structure of the 2LS-EMF system, paying special attention to its
relaxation and decoherence properties
\cite{Unr95,PSE,ZanRas,PleKni,SchMil,VioLlo,Gar,Zan}.

Environment-induced decoherence \cite{envdecrev} has been studied
extensively in recent years primarily based on models of quantum
Brownian motion (QBM)
\cite{FeyVer,CalLeg,GSI,UnrZur,HPZ,HM,HalYu,SolnHPZ} for the
interaction of a simple harmonic oscillator (Brownian particle)
with a harmonic oscillator bath (HOB) at a finite temperature,
leading to a reasonably good understanding of its characteristic
features. Decoherence of a 2LS in an EMF has been studied by a
number of authors, notably \cite{Unr95,PSE,VioLlo}, and their
dissipative and decoherent behavior are reported to be similar to
that of a QBM in a harmonic oscillator bath. They described the
progression in three stages --  quiescent, vacuum
fluctuation-dominated and thermal fluctuation-dominated,
separated  by the cutoff frequency and the thermal de Broglie
frequency (wavelength), which are characteristic QBM features
\cite{CalLeg,UnrZur,HPZ,PHZ,ZHP,HZ,AndHal}. A recent
comprehensive review of environment-induced decoherence can be
found in \cite{PZlh}.


Our findings, based on the standard 2LA-EMF model \cite{WM},  are
in stark disagreement from that reported in the literature. From
solutions of the exact master equation for the reduced density
matrix obtained recently by Charis Anastopoulos and the author
\cite{AH} capable of treating non-Markovian dynamics at zero
temperature, the decoherence time of the 2LA is found to be close
to the relaxation time. In many circumstances these times are
rather long -- indeed, that is why observations of many optical
coherence phenomena are possible. This result is, in first
appearance, rather counter-intuitive, and different from all
previous findings. However, as will be explained here, the
`intuition' one forms about the nature of dissipation and
decoherence are hitherto largely based on the QBM model. It is
quite general, yet by no means universal. This convenient
extrapolation could have influenced the choice of model in
previous authors' investigation and the readers' impression of
decoherence for a 2LS. Our findings show that such a commonly
invoked intuition for QBM in a HOB fails to apply to that of a
two-level atom (2LA) interacting with an electromagnetic field
(EMF). Our conclusion shows the need for extra caution in
accepting categorical depiction of decoherence: One needs to
identify the correct type of coupling of the 2LS with its
environment and consider  different decoherent mechanisms at
work in a realistic experiment.\\

\noindent {\it Decoherence in QBM}\\

The folklore that decoherence of  a Brownian oscillator proceeds
in a very short time (typically $10^{-40}$ of the dissipation
time) is really based on  high temperature ohmic bath conditions.
This is widely known because it is the only technically simple
case studied in detail \cite{CalLeg,envdecrev}.   It is by no
means universal. Intuitively, the bath needs to have many degrees
of freedom, preferably acting independently of each other so
incooperatively that the phase information in the system will be
dispersed to the largest extent amongst the many bath degrees of
freedom and affords little chance or takes inordinately long time
to be revived or reconstituted (recoherence, see, e.g.,
\cite{recoh}). Long decoherence times can appear in cases of low
temperature, supraohmic bath, as was first pointed out by Hu, Paz
and Zhang \cite{HPZ}in analyzing decoherence behavior of QBM based
on their exact non-Markovian master (HPZ) equation for general
environments. (For details based on numerical and analytical
solutions of the HPZ equation, see \cite{PHZ,SolnHPZ}). The almost
opposite picture (of very long decoherence time) is exemplified
by  two coupled subsystems where no coarse-graining is
introduced, such as the `dressed atom' description \cite{CPP} made
possible because of the coherence established between a 2LA and a
single resonant mode in the cavity. (Collapse and revival of the
Rabi nutation are distinct features of quantum coherence
\cite{WM}.) This shows that both the coupling and the nature of
the environment are important factors affecting the decoherence
of the system.

We also want to point out that equating a 2LS with QBM could be
as mistaken as ignoring the basic difference between dissipative
and mixing systems. To the extent that QBM exemplifies the former,
spin echo phenomena (e.g, Chap. 3 \cite{Wei})is an example of the
latter. Plasma waves showing  Landau damping in Vlasov dynamics
arising from a mean field approximation is another example \cite{HuPav,HKMP}
of the latter -- the `damping' is really a misnomer.
Just as in  the spin echo phenomena, the basic physics in
this  case is not  dissipation in  the  Boltzmann sense,
but  statistical mixing \cite{Ma}.  We will see that the
statistical mechanical properties of a 2LA-EMF system is
closer to the latter than the QBM, and reflects in their
different decoherence nature. \\

\noindent {\it Coherence in the 2LS}\\

For the 2LA-EMF system, one clear distinction between an EM field and a
system of harmonic oscillators acting as  bath is that the field (coupled to a detector)
has an intrinsic spectral density function, which cannot be chosen arbitrarily.
For example, it has been shown \cite{HM}
that a conformal scalar field in two dimensions coupled to a monopole detector
has an  Ohmic character while in four dimensions it is supraohmic .
 Barone and Caldeira \cite{CalBar} showed  that the spectral density function
for EM fields with momentum coupling to an oscillator detector is
supraohmic. These density functions would show very different
decoherent behavior from the high temperature Ohmic HOB case which
molds the common impression of decoherence.

But the most important  distinction  from QBM is that the 2LA
couples with the EMF in the discrete number basis for the field,
unlike the continuous amplitude basis in the QBM. This fact
(which is true in the rotating wave and dipole approximation)
implies that the 2LS plus EMF system is a {\it resonant} one.
Hence even though the EM field has just as many (in fact, a large
number of)  modes as the HOB,  only a very small fraction of them
in a narrow  range of the resonance frequency are efficiently
coupled to the atom. This is the root cause for the very different
qualitative behavior between the QBM and the 2LS as far as
decoherence is concerned.

We see that in a 2LA, for purely radiative decay the decay time
$T_1$ of the inversion is half  the decay time $T_2$ of the
polarization. There is no large order of magnitude differences
between dissipation and decoherence time. In fact it is perhaps
inappropriate to talk about dissipation for a 2LA-EMF system
because the conditions for a bath to actuate such a process is
lacking. The transition from excited to ground state is closer in
nature to relaxation (with relaxation time constant  $\Gamma$)
than dissipation. In a cavity where excitation of the atom from
the field (absorption)balances  with  emission, it is more
appropriate to refer to the  resonant state of the atom-field as
a coherent system. In these scenarios the distinction between QBM
and 2LA cannot be clearer. \\


\noindent {\it Difference between QBM and 2LS}\\

So what led earlier authors to make the claim that  2LS decoheres
easily? We think the confusion  arises when the picture of  QBM
dissipation and decoherence is grafted on the 2LA-EMF system
indiscriminately. If the field which acts as the environment is a
phonon field (from ion vibrations, see, e.g., \cite{Gar}), if
there is atomic collisions in a cavity \cite{Car} or if the
cavity walls are imperfect, some of these decoherent agents could
follow the QBM pattern as reported by many authors. Such sources
can be important for some setups.

Quantitatively, the model for the 2LS used by most authors for
the discussion of decoherence  inspired by QBM type of behavior
has the atom in a $\hat \sigma_z$  state (the diagonal Pauli
matrix) coupled to the field mode operators $\hat b^\dagger, \hat
b$. This type of coupling term (call it $\hat \sigma_z$ type for
convenience)   commutes with the Hamiltonian of the system, and
admits a diagonalization in the eigenbasis of the Hamiltonian.
The field is coupled to the atom as a whole, not to the two level
system (qubit), which is at the heart of quantum computation. By
contrast the standard model for 2LA-EMF which we studied has a
$\hat \sigma_\pm$ coupling to the field modes which controls the
two- level activity of the atom. This  coupling considered in the
standard model is indispensable,  i.e., it {\it cannot}  be
removed from the two-level atom as it {\it defines} it and will
be present in any realistic situation. Where then could the QBM
type of interaction enter in the 2LA?

If  the EM field is the only environment present,  we can still
ask if a QBM type of coupling term with the EM field would appear,
and if yes,  how strong would its effect be?  A useful way to
compare the relative importance of these two types of coupling
would be to seek out the source where they stem from.

Recall that the standard model is derived  under the dipole and
rotating wave approximations.
 In the next section we will show that the
$\hat \sigma_z$ type of coupling appears only in the next order
expansion after the dipole approximation. Since these are good
approximations for a large class of atomic states when the atom
is nonrelativistic, the contribution from the QBM type of
coupling used in  \cite{Unr95,PSE,ZanRas,VioLlo} should be
negligible in a 2LA-EMF system and its ensuing decoherent effect
insignificant. In this sense the EM field does not in leading
order of approximation act like a bath in the QBM way,  and
coherence  in a 2LA-EMF system  is quite well preserved
(excepting other processes, e.g. \cite{SchMil,PleKni}).

Decoherent behavior of the 2LA-EMF system we are reporting on
here is based on an analysis of the full (non-Markovian) dynamics
in the cases of a free quantum field and a cavity field at zero
temperature from solutions of an exact master equations derived
by Anastopoulos and Hu \cite{AH}. They used the influence
functional method \cite{FeyVer} to take into account the full
backreaction of the field on the atom, while adopting
Grassmannian variables for the 2LA and the coherent state
representation for the EMF.


\section{The Model}

Our model for atom-field interaction is the standard one (see
Appendix A of \cite{AH} for details)  \cite{WM,MW,Wei}
\footnote{Our Hamiltonian is given in the so-called minimal
coupling (MC) as different from the multipolar coupling (MP)
\cite{CPP}, which may be more relevant to atoms in a cavity
because the explicit Coulomb interaction between the atom and its
image charge is removed.}. The total Hamiltonian for a
(stationary) atom interacting with a quantum electromagnetic
field (EMF) under the dipole, rotating wave (RW) and two-level
(2L) approximation is given by

\be {\hat H} = \hbar \o_0 {\hat S}_z + \hbar \sum_\bk \left[
\o_\bk {\hat b}_\bk^{\dagger} {\hat b}_\bk +    \left(g_\bk  S_+
{\hat b}_\bk  +\bar g_\bk  S_- {\hat  b}_\bk^{\dagger} \right)
\right] \label{II.1} \ee where ${\hat b}_\bk^{\dagger},{\hat
b}_\bk$ are the creation and annihilation operators for the kth
normal mode with frequency $\o_\bk$ of the electromagnetic field
(thus for the field vacuum ${\hat b}_\bk|0\rangle = 0, [{\hat
b}_\bk, {\hat b}_{\bk'}^{\dagger}] = \delta_{\bk,\bk'}$,
 for all $\bk$.), and $ \o_0= \o_{21}$ is the frequency between the two levels.
Here \be \hat S_z = \ha \hat \s_z, \hat S_\pm = \hat \s_\pm
\equiv \ha (\hat \s_x \pm i \hat \s_y) \ee
 where $\hat \s_{x,y,z}$ are
the standard 2x2 Pauli matrices with $\hat \s_z = diag (1, -1)$,
etc. The coupling constant $g_\bk \equiv  d_{21\bk} f_\bk (\bX)$
where
\begin{equation}
d_{ij\bk} \equiv -\frac{i\o_{ij}}{\sqrt{2\hbar\o_\bk\epsilon_0 V}}\bd_{ij}
\cdot \hat {\bf e}_{\bk \s}
\end{equation}
and  $\bd_{ij} \equiv e \int {\bar \phi_i}{\bf x}\phi_j d^3x$ is the dipole
matrix element between the eigenfunctions $\phi_i$ of the electron-field
system,
 $\hat {\bf e}_{\bk \s}$ is the unit polarization vector ( $\s =1, 2$ are the
two polarizations), and $f _\bk (\bx)$ is the spatial mode
functions of the vector potential of the electromagnetic field
(in free space, $f_\bk (\bx) = e^{-i \bk \cdot \bx}$,  $V$ is the
volume of space.). Under the dipole approximation $f_\bk$ is
evaluated at the position of the atom $\bX$. Since $ \bd_{ij} =
{\bar \bd}_{ji}$, ${\bar d}_{ij\bk} = d_{ji\bk}$, we will choose
a mode function representation such that $g_\bk$ is real. When
only one mode in the EM field is considered, this is the
Jaynes-Cummings model.

To see how this could possibly be related to the $\hat \sigma_z$
type of coupling with Hamiltonian (used by e.g.,
\cite{PSE,VioLlo} for the study of decoherence in 2LS) \be {\hat
H} = \hbar \o_0 {\hat S}_z + \hbar \sum_\bk \left[ \o_\bk {\hat
b}_\bk^{\dagger} {\hat b}_\bk + \hbar \hat \sigma_z   \left(\bar
g_\bk {\hat b}_\bk  + g_\bk  {\hat  b}_\bk^{\dagger} \right)
\right] \label{II.3} \ee we examine the next term after the
dipole approximation. This has a contribution to $g_{ij \bk}$
even when $i=j$. This is  equal to
\begin{equation}
g_{ii {\bf k}} = c_{{\bf k}} {\bf k} \cdot {\bf q}_i
\end{equation}
where
\begin{equation}
{\bf q}_i = \sum_{ \sigma} \int \bar{\phi}_i {\bf \delta x} ({\bf
p} \cdot {\bf \hat{e}}_{{\bf k} \sigma} ) \phi_i dx^3
\end{equation}
and $c_k$ is a constant given by
\begin{equation}
c_{\bf k} = - \frac{e}{m} (2 \hbar  \omega_{{\bf k}} \epsilon_0 V)^{-1/2}
\end{equation}
It generates  an additional coupling term
\begin{equation}
       \sum_{\bf k} \hat \sigma_z ( g_{1{\bf k}} b_{{\bf k}} +\bar  g_{1 {\bf k}}
b^{\dagger}_{{\bf k}}) + 1 (g_{2{\bf k}} b_{{\bf k}} +\bar  g_{2 {\bf k}}
b^{\dagger}_{{\bf k}})
\end{equation}
where
\begin{equation}
g_{1 {\bf k}} = g_{11{\bf k}} - g_{22 {\bf k}}, \,\,\,
g_{2 {\bf k}} = g_{11{\bf k}}+ g_{22 {\bf k}}
\end{equation}
This  gives  the lowest order  $\hat \sigma_z$ type of coupling
in a 2LA -EMF system. The ratio of  the coupling $g_{1{\bf k}}$
of the $\hat \sigma_z$ type in (\ref{II.3})to the dipole coupling
$g_\bk$ in (\ref{II.1}) is
\begin{equation}
|g_{1 {\bf k}}/g_{ {\bf k}}| = | \frac{{\bf k} ({\bf q}_1 - {\bf q}_2)}{ m
\omega_\bk  d_{12}}| \leq \frac{\omega_\bk |{\bf q}_1 -{\bf q}_2|}{m \omega_\bk d_{12}}
\end{equation}
Thus the $\hat \sigma_z$ type of coupling generated from the 2LA-
EMF interaction will be  significant only for very  high
frequencies $\omega_{\bf k}$ of the EM field, a  point intuitively
clear from the meaning of the dipole approximation.

\section{The Master Equation}

We use the open system approach and the influence functional
formalism to derive a master equation for the 2LA with
backreaction of the field treated self-consistently. This
involves deriving the influence functional and  the evolution
operator for the reduced density matrix of the 2LA.  For
convenience we use Grassmannian variables for treating fermions,
and the coherent state representation for the field. The coherent
state of the combined atom-field system is
\begin{equation}
|\{z\},\eta  \rangle =|\{z\}\rangle \times |\eta  \rangle
\end{equation}
where  $|z\rangle $, $z$ a complex number, denotes the EM field
coherent states and $|\eta\rangle$, $\eta$ a Grassmannian or
anticommuting number, denotes the electron coherent state. We
assume initially that the density matrix of the total
system+environment is factorizable $ \hat \rho (0) = \hat \rho_e
(0) \otimes \hat \rho_b (0)$. Only at that time would $z$ and
$\eta$ be pure complex and Grassmannian numbers respectively. As
the system evolves, both $\eta  $ and $z$ contain Grassmann and
c-number parts. The mixing of even and odd parts (note $g_{\bf k}$
is odd) comes about as the initially factorized atom state becomes
"dressed".

We skip over the details of the derivation which can be found in
the original paper \cite{AH} but simply present the master
equation, here in operator form (at zero temperature with the
field in a vacuum state) as follows: Writing
 \begin{equation}
\frac{\dot{u}(t)}{u(t)} = \Gamma (t) +i \Omega(t)
\end{equation}
the master equation reads
\begin{equation}
\frac{\partial}{\partial t} \rho = -i  [H(t) ,\rho] + \Gamma (t)
\{S_+S_-,\rho \} - 2 \Gamma (t) S_- \rho S_+ \label{III.17}
\end{equation}
where
\begin{eqnarray}
H(t) = \Omega(t) S_+ S_-
\end{eqnarray}
The first term corresponds to the unitary Hamiltonian evolution, only now
the effect of the environment has induced a time dependent shift in the
value of the frequency, the second term is time dependent dissipation and
the third corresponds to noise.

The effect of the field is contained in the $u(s), \bar u(s)$
functions, which are obtained as solutions of the linear
integro-differential equations
\begin{eqnarray}
\dot{\eta }+i\omega \eta  +\int_{0}^{s}ds^{\prime }\mu  (s-s^{\prime })\eta
(s^{\prime }) &=&0\hspace{4cm} \\
\dot{\bar{\eta }^{\prime }}-i\omega \bar{\eta }^{\prime
}+\int_{0}^{s}ds^{\prime }\mu  ^{*}(s-s^{\prime })\bar{\eta }^{\prime
}(s^{\prime }) &=&0\hspace{4cm}
\end{eqnarray}
under the condition
\be
u(0)=\bar{u}(t)=1
\ee
Here, the kernel $\mu(s)$ is given by
\begin{equation}
\mu  (s)=\sum_{\bf k}  g_{\bf k}^{2}e^{-i\omega _{\bf k}  s}
\end{equation}
The  equations for $u(t)$ functions can be solved with the use of
the Laplace transform and the  convolution theorem. They are given
by
\begin{equation}
u(s)={\cal L}^{-1}\left( \frac{1}{z+i\omega +\tilde{\mu }(z)}\right)
=\frac{1%
}{2\pi i}\int_{c-i\infty }^{c+i\infty }\frac{dze^{zs}}{z+i\omega +\tilde{\mu
}(z)}
\end{equation}
where $\tilde{\mu }(z)$ is the Laplace transform of the kernel
$\mu(s)$ and $c$ is a real constant larger than the real part of
the poles of the integrand.

\subsection{Spontaneous emission}

To show how the standard results are regained, and to understand the
meaning of the new function in the master equation, let us consider
the physical process of spontaneous emission.  Start with a generic
initial density matrix
\be
\rho =\left(
\begin{array}{cl}
1-x & y \\
y^* & x
\end{array}
\right)
\ee
its corresponding Q-symbol is
\begin{equation}
\rho (\bar{\eta },\eta  )=x+y^{*}\eta  +y\bar{\eta }+(1-x)\bar{\eta }\eta
\end{equation}
If we evolve it with the density matrix propagator (derived in
\cite{AH}) we obtain for the state at time $t$
\begin{equation}
\rho _{t}(\bar{\eta },\eta  )= 1 - \bar{u}u(1-x)
+\left(
\bar{u}%
y^{*}\eta  \right) +\left( u\bar{\eta }y\right) +\left( \bar{u}u(1-x)\right) \bar{\eta} \eta
\end{equation}
corresponding to
\be
\rho _{t}=\left(
\begin{array}{cl}
\bar{u}u(1-x) & uy \\
\bar{u}y^* & 1-\bar{u}u(1-x)
\end{array}
\right) \ee Considering the case $x=y=0$ we get for the
probability of spontaneous emission \be P(1\rightarrow
0,t)=1-\bar{u}u \ee The rate of decoherence in the energy
eigenstates is governed by the absolute value of the function $u$
(the off- diagonal terms) while $u$ itself determines the rate of
energy flow from the atom to the environment. Hence for our
particular choice of initial state (vacuum) we find that
decoherence and relaxation time are essentially identical . This
equation is useful for studying decoherence of a qubit in a QED
cavity.

\subsection{Field modes in Free Space and Cavity }

Our master equation (\ref{III.17}) depends solely on the function
$u(t)$, which in turn is determined by the kernel $\mu (s)$. In
\cite{AH} we have given some analytic expressions for this
function in various cases, including a single mode, an infinite
number of modes in the field, and a cavity consisting of two
parallel plates at distance $L$. The field satisfies Dirichlet
boundary conditions on the surface of the plates. In all cases
our  results obtained from the generalized master equation bear
closer resemblance qualitatively to the 2LA behavior than that of
quantum Brownian motion. Details can be found in \cite{AH}. We
now summarize the major findings obtained there.

\section{Discussion}

The physics of a 2LA-EMF system at zero temperature
is characterized by a number of time constants:\\
1) The inverse natural frequency $  \omega_0^{-1}$ \\
2) The inverse coupling constant $ g_{\bf k}^{-1}=\sqrt {\omega_{\bf k}}/\lambda$\\
3) The relaxation time constant $  \Gamma^{-1} $\\
4) The cavity size $ L $  (divided by c)

First consider a zero temperature field in free space, thus
ignoring factor 4). Start with only one mode in the field in
resonance with the atom (whence we omit the subscript in $g_{\bf
k}$), then the system undergoes Rabi nutation with frequency
$\Omega \approx g \sqrt{n+1}$, where $n$ is the  photon number in
the field. The collapse time (assuming a large mean  photon number
$\bar n$) is $~g^{-1} $,
 and revival time is $~ 2 \pi \sqrt{\bar n}/g$. \cite{WM}.
Atom excitation becomes significant in a time much greater than  $
\omega_0^{-1}$ but shorter than $g^{-1}$.
(This  is the condition for a first order perturbation
theory to give reasonable results.)
For a large number of modes, spontaneous emission occurs at the
relaxation time scale  $\Gamma^{-1} = \pi /g >> \omega_0^{-1}$
which we found  to be the same as the decoherence time -- the
time for the off diagonal elements  of the reduced density matrix
to decay. When the mean  number of photons  in the field is large
($\bar n >> 1$), they become comparable to the collapse time. This
is a measure of the coherence in the atom-field system, and is
controlled mainly by their coupling and the photon number in the
field.  We see that with the resonance condition, the  nature of
decoherence in 2LS   is very different from the QBM situation,
where phase information in the Brownian particle is efficiently
dispersed in the many modes in the bath coupled almost equally to
the system. As we remarked in the Introduction, the
identification of the phase information and energy flow from the
2LS to its environment is similar to the spin echo phenomena
(Landau `damping') which is based on statistical  mixing rather
than dissipation.  The mathematical distinction lies between
considering the system coupled to the continuous amplitude basis
(QBM) of the environment or to the discrete number basis (2LA).
The QBM case essentially produces noise that drives the system in
a way insensitive to its own intrinsic dynamics. In our model, the
coupling respects the internal dynamical structure of the 2LS and
allows it to keep its coherence.

To see how the distribution of modes in a field changes the
picture,  the cavity field calculation is useful. There, as the
plots in \cite{AH} show, the relaxation constant develops peaks
and minima. The resonance effect is enhanced by a cavity size
commensurate with the natural frequency of the 2LA and
dissipation  weakens. Narrow band resonance fluorescence as well
as inhibition of spontaneous decay by frequent measurements --
the Quantum Zeno effect -- are interesting phenomena  which our
equations can provide finer details.

Non-Markovian processes involve memory effects (nonlocal in
time). When the reaction time of the bath is comparable to or
faster than the natural time scale of the system ($\omega_0$),
one also expects to see non-Markovian behavior. For the QBM
problem, the case of high temperature Ohmic bath is almost the
only condition that would yield a Markovian dynamics. For other
types of spectral density (supraohmic) or for a low temperature
bath, the dynamics of the system is generally non-Markovian
\cite{HPZ}.  By contrast, the 2LA is quite different: At zero
temperature there is only one time scale $\Gamma^{-1} =
\lambda^{-2} \omega^{-1} >> \omega^{-1}$ that determines both
decoherence and relaxation. There is no memory effect and hence
the process is Markovian. We expect that at finite temperature
the dynamics of the 2LA will be nonMarkovian \cite{ADHS}. This is
because there are more ways for  the atom and the field to get
entangled, and the memory effects of their interaction would
presumably persist.


To conclude, we show that not all two level systems (2LS) decohere
like quantum Brownian particles (QBM). Specifically, in the case
of a two-level atom (2LA) interacting with an electromagnetic
field (EMF) alone through the standard $\hat \sigma_{\pm}$
coupling as in a quantum optics / cavity QED quantum computer
prototype, decoherence time of the 2LA is close to the relaxation
time. This behavior is completely different from the QBM. The
crucial difference lies in the type of coupling between the 2LS
and the environment: The 2LA interacts resonately with selected
modes in the electromagnetic field while the QBM is coupled to
all modes in the harmonic oscillator bath. Only for those
mechanisms which can be described by a $\hat \sigma_z$ type of
coupling of the 2LS with the environment (which is subdominant in
a 2LA-EMF system) will one expect a QBM-like decoherent behavior.
Therefore one has to be careful in specifying what type of 2LS
constitutes the qubit and and how it couples to the field, since
the field can act as a resonator (2LA) or as a bath (QBM)
producing very different decoherent behavior. In a realistic
experiment one needs to know  which processes are describable by
what type of coupling and the weight of their relative
contributions in varying environmental conditions before drawing
credible conclusions about the overall nature and degree of
decoherence in that system. \\

{\bf Acknowledgement} This is a summary of work done with Dr.
Charis Anastopoulos based on \cite{AH}, and ongoing work with
Adrian Dragulescu and Sanjiv Shresta, with whom I have enjoyed
many useful discussions and fruitful collaboration. This work is
supported in part by NSF grant PHY98-00967.


\end{document}